\documentclass{eas}
\usepackage{graphicx}
\begin{document}

\TitreGlobal{Mass Profiles and Shapes of Cosmological Structures}

\title{Elliptical Galaxy Halo Masses \\from Internal Kinematics}
\author{Romanowsky, A.J.}\address{Departamento de F\'{i}sica, Universidad de Concepci\'{o}n, Casilla 160-C, Concepci\'{o}n, Chile}
%
\runningtitle{Elliptical Galaxy Halo Masses from Internal Kinematics}
\setcounter{page}{1}
\index{Romanowsky, A.J.}

%
\begin{abstract}
The halo masses of nearby individual elliptical galaxies
can be estimated by using the kinematics of their stars, planetary nebulae, and globular
clusters---ideally in combination.
With currently improving coverage of galaxies of ordinary luminosities and morphologies,
systematic trends may be identified.
Bright, boxy ellipticals show strong signatures of dark matter,
while faint, disky ones typically do not.
The former result is problematic for the MOND theory of gravity, 
and the latter is a challenge to explain in the
$\Lambda$CDM paradigm of galaxy formation.
\end{abstract}
\maketitle
%
\section{Introduction}

While the mass profiles of spiral galaxies can be studied
via their extended cold gas disks, elliptical galaxies
rarely offer this avenue.  An obvious alternative
is to measure the kinematics of the integrated
stellar light, but this is observationally prohibitive in
the low surface brightness outer regions where the mass profile
is of the most interest.
Even when the kinematics are measured, interpretation is
more difficult than in spirals because
of uncertainties in the intrinsic shapes,
viewing angles, and orbit types.
Additional complications are dust effects (Baes \& Dejonghe 2001)
and stellar population bias (De Bruyne et al. 2004).

It is possible to surmount many of the obstacles to
interpretation with sufficient data and sophisticated models.
The orbit types can be determined using
higher order moments of the velocity distribution
(van der Marel \& Franx 1993),
and the shapes and viewing angles can be constrained using
integral field data (Cappellari et al. 2005 = C+05).
It remains to be seen if all the degeneracies can be removed
when only 3 of the 6 phase space dimensions are probed.
In any case, there has until now been no systematic survey of
individual nearby elliptical galaxies which includes an unbiased sample,
extended data, and adequate models---and thus general conclusions about
their dark matter halos should be considered tentative.

\section{Results from integrated stellar kinematics}

While there have been many studies of stellar kinematics in ellipticals,
so far the most reliable survey for dark matter (DM)
was from Kronawitter et al. (2000) and Gerhard et al. (2001 = G+01).
With long-slit data on 21 bright, round ellipticals,
they found that the circular velocity $v_{\rm c}(r)$ is
roughly constant
to 1--2 $R_{\rm eff}$ (5--10 kpc; see Fig.~1, left).
Statistically, the central DM
density appeared to be 25 times larger in ellipticals than in spirals.
But on an individual basis,
a constant mass-to-light ratio $(M/L)$ was ruled out for only 5 of the brighter galaxies,
so these results are not broadly conclusive.
There are a couple of other galaxies
where extended stellar kinematics indicate a massive dark halo
(Statler et al. 1999; Thomas et al. 2005).
Note that with $\Lambda$CDM halos around elliptical galaxies,
there should be no convenient ``flat'' part of the rotation curve
where one can make a uniform Tully-Fisher measurement
(as assumed in studies such as Ferrarese 2002).

\begin{figure}[t]
   \centering
   \includegraphics[width=6.195cm]{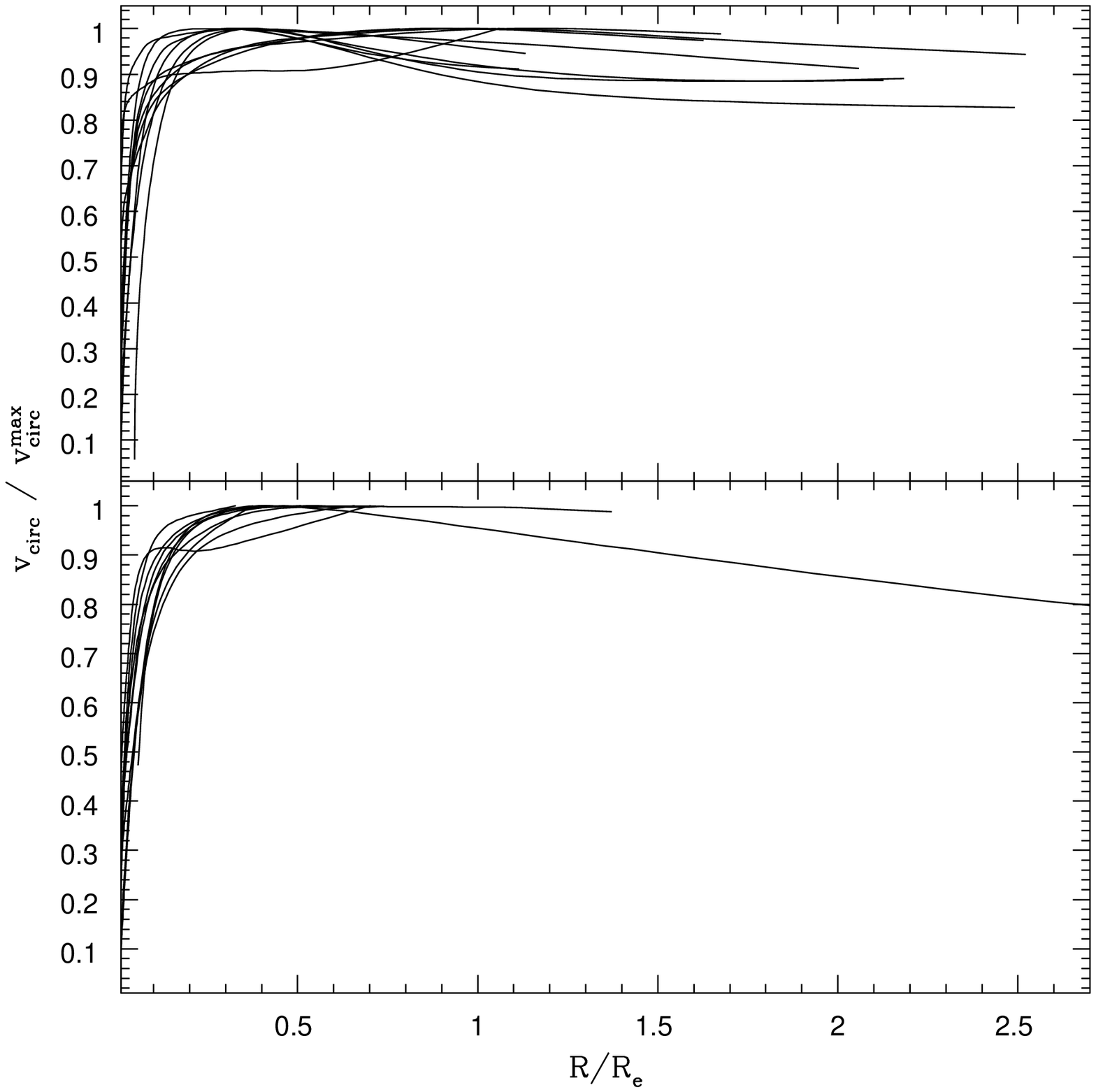}
   \includegraphics[width=6.195cm]{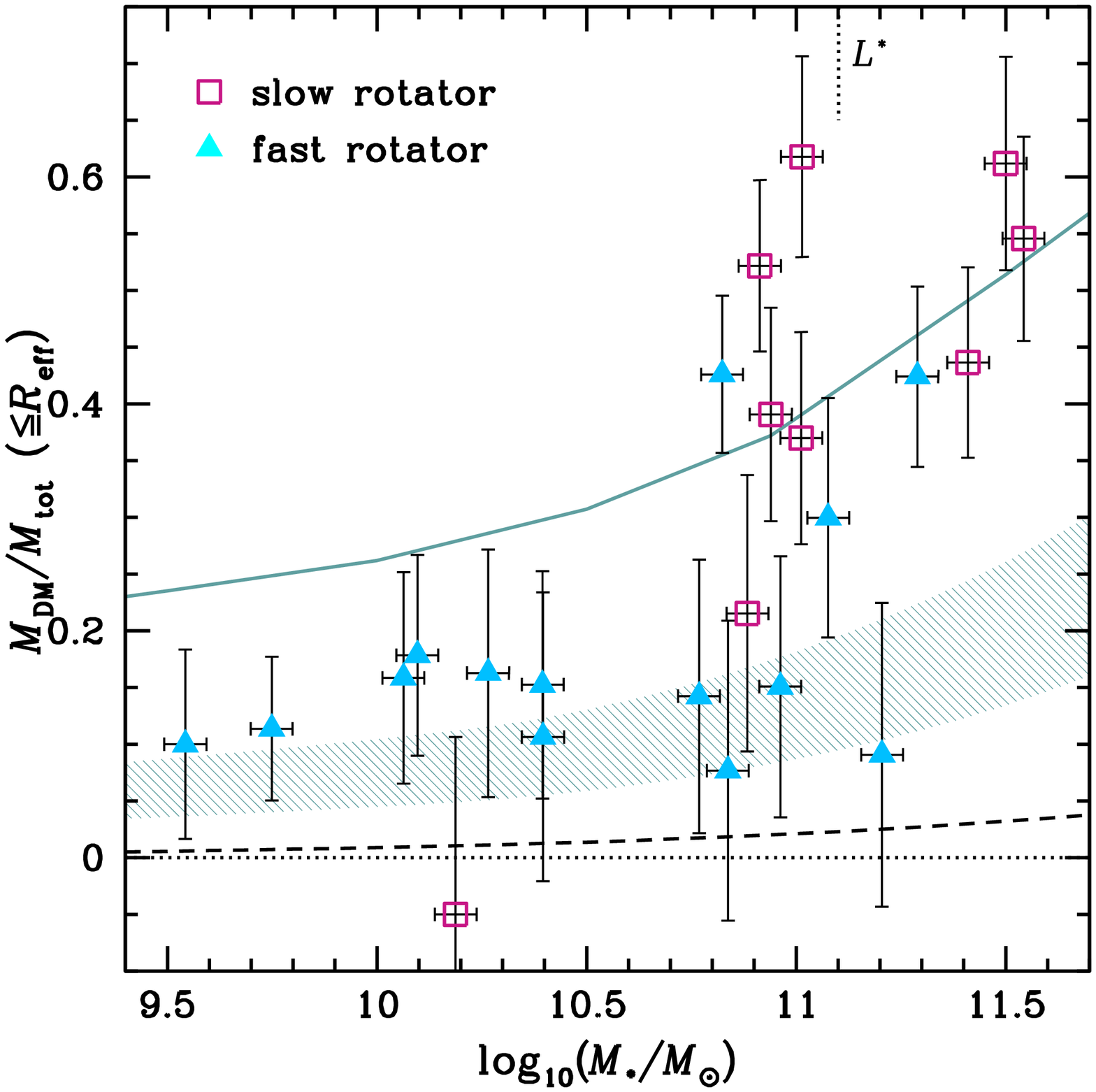}
      \caption{ 
	{\it Left:} Rescaled circular velocity profiles of
	elliptical galaxy sample from G+01.
	{\it Right:} Central dark matter fraction in SAURON
        sample of ellipticals.
	The shaded region shows the 1-$\sigma$ range with 
	dissipationless $\Lambda$CDM halos included.
	The solid line illustrates the effects of adiabatic contraction.
	The dashed line shows the MOND prediction.
	}
       \label{figure_mafig1}
   \end{figure}

In addition to G+01,
there have been other approaches to deciphering the central DM content
(Borriello et al. 2003; Padmanabhan et al. 2004;
Trujillo et al. 2004; Mamon \& {\L}okas 2005a; C+05).
Almost all of them infer that DM is not dominant within $R_{\rm eff}$,
comprising $\sim$~30\% of the mass as a typical estimate.
Of particular interest is C+05, whose SAURON
data and models are the best to date.
Given their independent mass estimates from dynamics and stellar populations,
we infer the DM fraction at a uniform physical radius, and compare
these to theoretical predictions (see Fig. 1, right).
The rotation-dominated galaxies are consistent with harboring 
dissipationless $\Lambda$CDM halos, while the anisotropy-dominated
galaxies follow a trend expected for dissipative halo contraction
(Blumenthal et al. 1986).
However, a small systematic error could imply that the fast rotators
have no DM at all, showing the limitations of working in the
central regions only.

Modified Newtonian Dynamics
(MOND) would explain these SAURON results with difficulty, requiring
large {\it ad hoc} systematic errors.
Also,
the onset of mass discrepancies in ellipticals appears at higher accelerations
than the universal MOND value (G+01).
However, the uncertainties in this analysis are unclear.

\section{Techniques and programs for large radius}

Mass tracers well into the galactic halos are clearly needed.
Two ubiquitous kinematical probes are planetary nebulae (PNe) and
globular clusters (GCs).
PNe are somewhat easier to observe in nearby galaxies,
and importantly provide contiguous constraints with the central
stellar kinematics.
GCs are observable at larger distances and more abundant at larger radii.
While $\sim$1000 discrete velocities are normally required
to fully constrain a hot dynamical system (Merritt \& Saha 1993),
in galaxies, many fewer are needed because of the 
additional strong constraints on the central regions provided by stellar kinematics.

There are so far a handful of galaxies with PN or GC kinematics
measured 
(most are referenced in Napolitano et al. 05 = N+05).
The largest PN study is of NGC~5128, with $\sim$800 velocities
(Peng et al. 2004). This peculiar galaxy shows evidence of a surprisingly weak DM
halo, with $M/L_B \sim$~13 inside 80 kpc.
The largest GC study is of NGC~1399, with $\sim$700 velocities to 90 kpc
(Richtler et al. 2004 and in prep).
The constant GC velocity dispersion implies a massive DM halo as
expected for this central Fornax Cluster galaxy.

Another halo probe is X-ray emitting hot gas.
With the advent of {\it Chandra} and {\it XMM-Newton},
it is now possible in some galaxies to 
remove contaminating point sources, 
check the equilibrium of the gas, and
determine the gas temperature profile
(Fukazawa et al. 2006).
However, the temptation of X-ray studies is to study the highest
$L_X$ systems, which gives a systematically biased picture of 
galaxy masses.  It is important to control the selection effects.

Given the challenges of determining mass profiles, it is ideal
to combine as many halo probes as possible, e.g. PNe, GCs, X-rays.
The first exercise is to cross-check these for reliability (see Fig. 2, left).
If they can all be used confidently, in combination the constraints
are much stronger.  E.g., one could derive the mass profile from
X-rays and determine the orbit structures from kinematical data.
Such a combined halo strategy is now underway. 
A long-term program with the PN.Spectrograph (PN.S; Douglas et al. 2002)
is studying the PN kinematics.  Various GC projects are underway at
CTIO, Gemini, Magellan, and the VLT, 
as well as X-ray projects (O'Sullivan et al.)
The framework of these surveys is to investigate the properties of
{\it ordinary} elliptical galaxies ($\sim L^*$), as a function of 
environment, and comparing the two 
different families of boxy and disky galaxies
(Kormendy \& Bender 1996).

Dynamical interpretations of elliptical galaxies now often use
the orbit modeling approach invented by Schwarzschild (1979)
and extended by many others (see Thomas et al. 2005).
Assuming a functional form for the gravitational potential,
one builds a comprehensive library of possible orbits.
The contribution of each orbit to each observable is calculated,
and the best-fitting orbit combination is found.
Such methods (see also G+01)
are physical by construction, and fully nonparametric
in their orbit solutions.
One can also make optimal use of the discrete velocities
using maximum likelihood 
(Fig.~2, left: Romanowsky \& Kochanek 2001; see also Wu \& Tremaine 2005).
Convergence issues can be addressed by
attention to library size and regularization bias
(Richstone et al. 2004).

\begin{figure}[t]
   \centering
   \includegraphics[width=6.195cm]{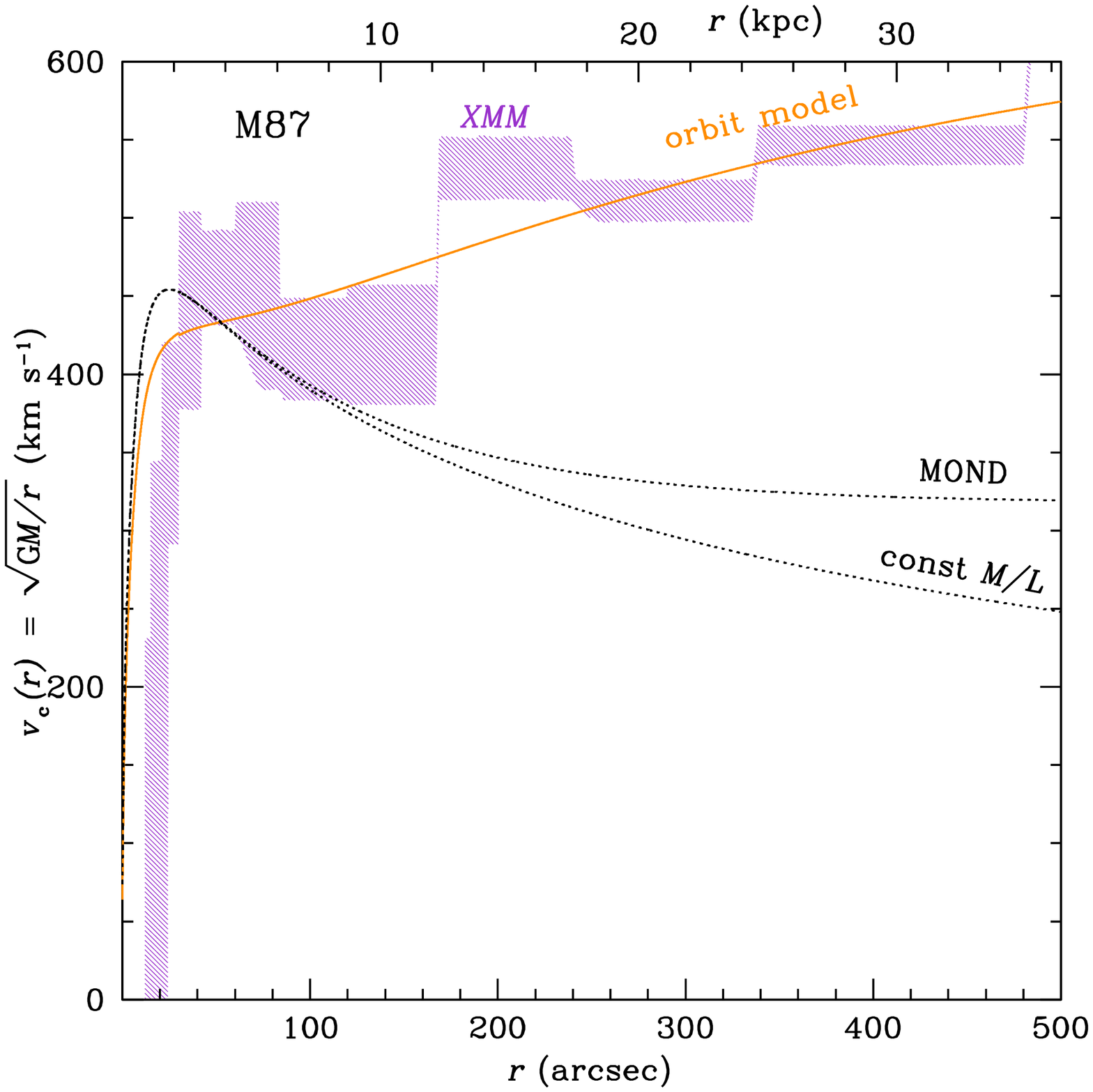}
   \includegraphics[width=6.195cm]{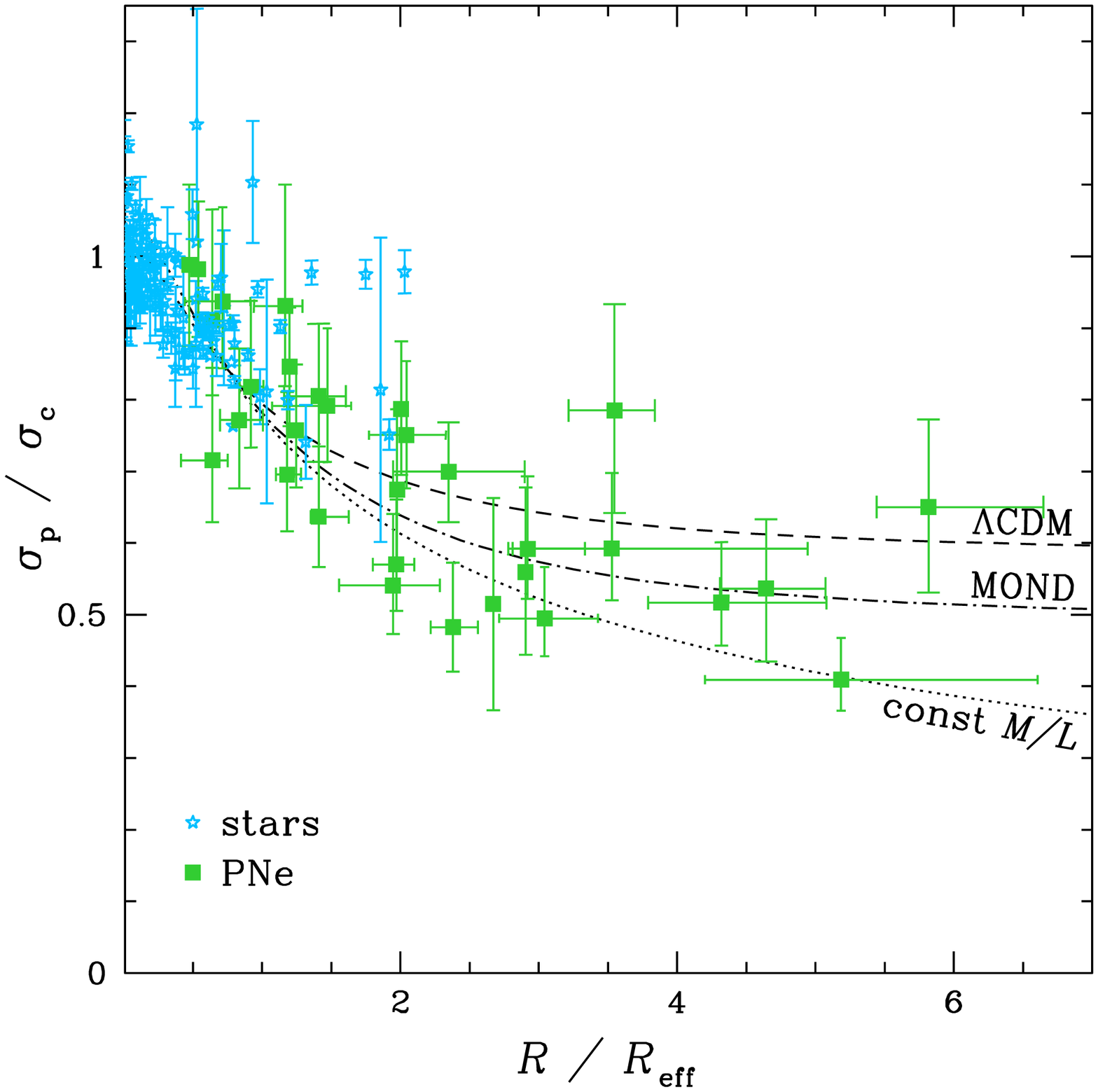}
      \caption{ 
	{\it Left:} Modeled circular velocity profile of M87.
	Anisotropic orbit model uses stellar kinematics + 234 GC velocities.
	{\it XMM-Newton} model is from Matsushita et al. (2002).
	Dashed curves are isotropic models, showing that
	the GCs are near-isotropic, while the stars+PNe are not.
	{\it Right:} Projected velocity dispersion profiles,
	rescaled and stacked, of stars and PNe in 5 elliptical 
	galaxies, compared to isotropic models.
	}
       \label{figure_mafig2}
   \end{figure}

\section{First halo constraints on ordinary ellipticals}

Five $L^*$ ellipticals now have their PN halo kinematics measured
(M\'{e}ndez et al. 2001; Romanowsky et al. 2003; Teodorescu et al. 2005).
Four of them show a similar decline in velocity dispersion with radius
(Fig. 2, right), which is also seen in the Milky Way halo (Battaglia et al. 2005).
While the decline is caused in part by the compact nature of the
ellipticals' stellar mass distribution,
it is still more extreme than expected from isotropic models with DM.
However, isotropy is an arbitrary assumption, and there is
an emerging consensus that radial orbits are expected in stellar halos
(S\'{a}iz et al. 2004; Dekel et al. 2005; Diemand et al. 2005; Abadi et al. 2005;
but see Gonz\'{a}lez-Garc\'{i}a \& Balcells 2005; Athanassoula 2005)---which could 
help produce the observed dispersions (Mamon \& {\L}okas 2005b).
More definitive mass determinations obviously require careful consideration of
the anisotropy, which we have done for NGC~3379 as the first case.

For NGC~3379, we used spherical orbit models, with the
constraints including stellar kinematics and 109 PN velocities.
The best-fit models have an anisotropy parameter $\beta$ varying
from $-0.3$ in the center to $+0.5$ in the outer parts.
The cumulative $M/L_B$ at $5 R_{\rm eff}$ is $7.6 \pm 0.9$,
implying a DM fraction of 15--30\% within this radius,
which is surprisingly low.
However, a typical $\Lambda$CDM halo is
still allowed within the uncertainties (see Fig. 3, left).
The situation may be clarified by the addition of the SAURON data and
$\sim$100 more PN velocities now obtained, along with refinements of the
modeling.

There are other available constraints on the NGC~3379 halo.
One is a HI-emitting gas ring in a near-Keplerian orbit at 100 kpc
(Schneider 1985).
The implied $M/L_B$ is $27\pm5$, startlingly low compared to
expectations of $\sim$150 for a group core (van den Bosch et al. 2003).
We have also obtained 38 GC velocities to distances of 40 kpc
(Bergond et al. 2006).  These show a flat dispersion profile,
indicating a more massive halo than the HI ring implies, and making
a consensus solution between HI, PNe, and GCs difficult
(see Fig. 3, left).
Anisotropic GC models (with more velocities) are
necessary to unravel this puzzle.
Note that the PN and GC data are feasible with MOND, but the
HI constraint is not.

\begin{figure}[t]
   \centering
   \includegraphics[width=6.195cm]{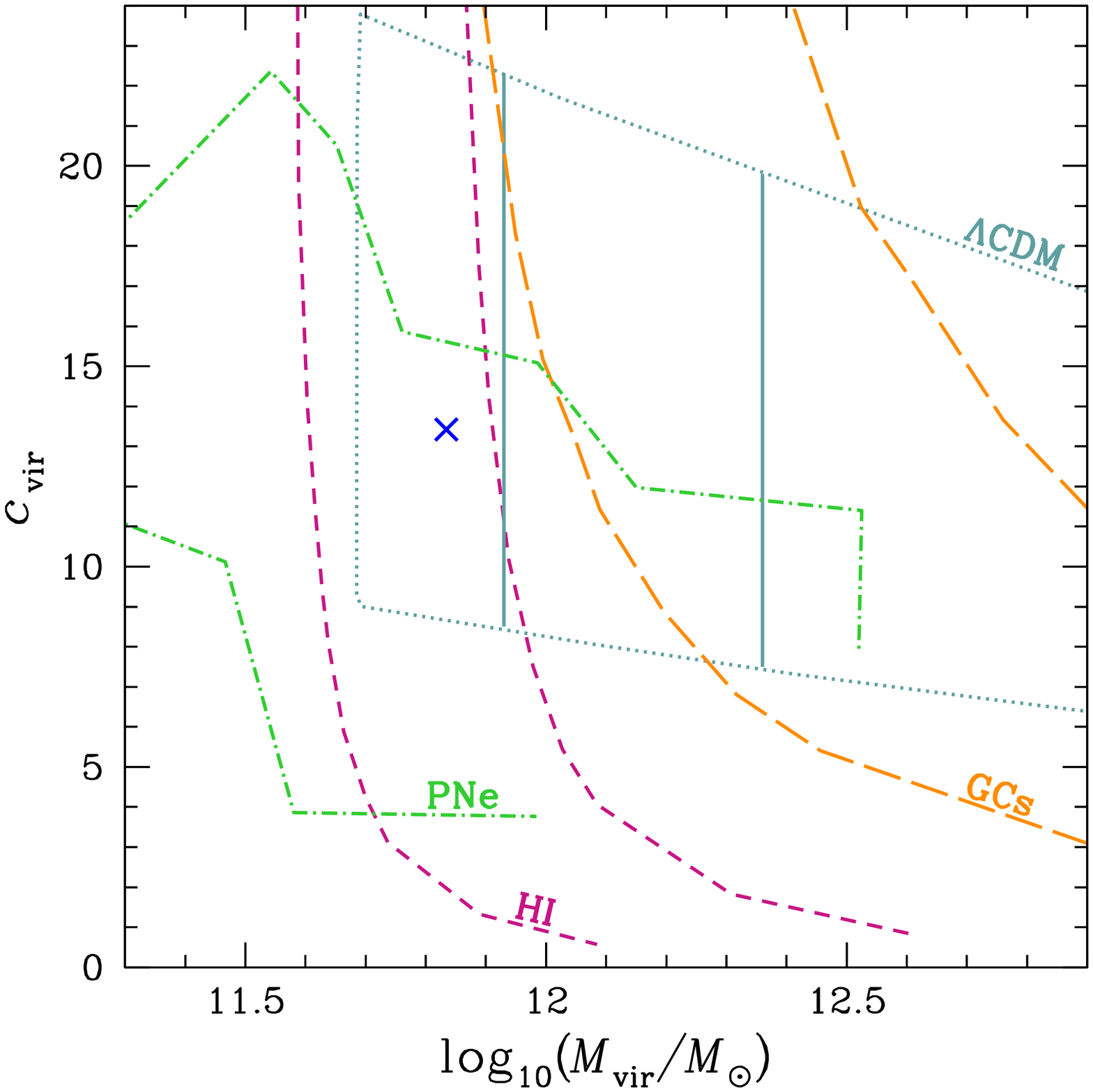}
   \includegraphics[width=6.195cm]{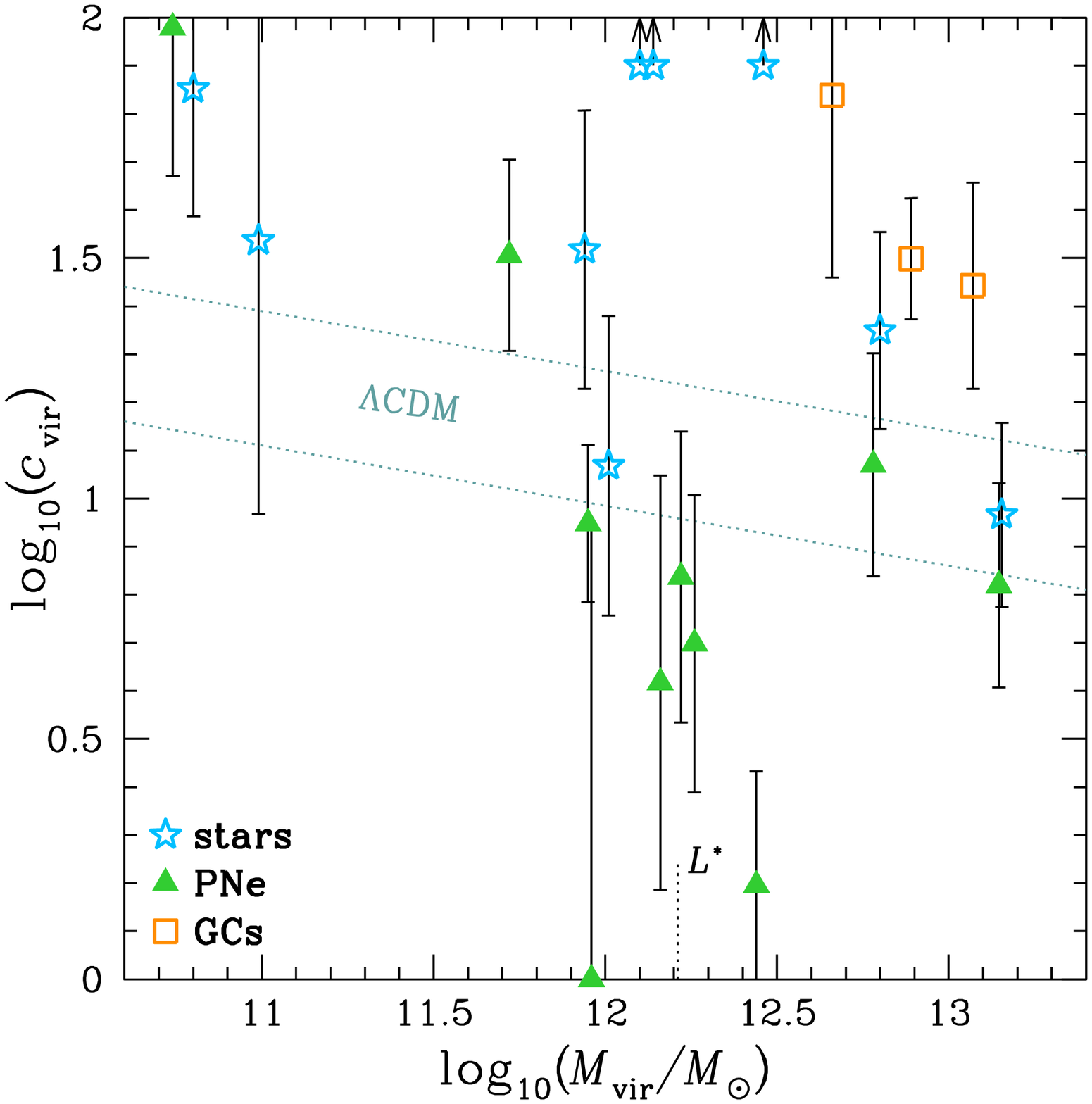}
      \caption{ 
	Constraints on halo masses and concentrations.
	{\it Left:} NGC 3379, with various constraints shown
	as contours of 68\%.
	For the $\Lambda$CDM boundaries, solid lines show
	predictions for a group or galaxy halo (van den Bosch et al. 2003),
	while dotted lines show the full range of physical plausibility.	
	The best PNe+GCs+HI+$\Lambda$CDM solution is marked by $\times$.
	{\it Right}: Results from a sample of early-type galaxies (after N+05).
	}
       \label{figure_mafig3}
   \end{figure}

\section{Implications for $\Lambda$CDM and MOND}

While firm results await full analyses of the PN.S (etc.) samples,
in the meantime we can examine the available kinematical evidence on
mass profiles.  We have assembled the literature data on
the masses of ellipticals at $\geq 2 R_{\rm eff}$ (N+05).
Assuming dissipationless $\Lambda$CDM halos, we make simple model fits for
 virial masses and concentrations (see Fig. 3, right).
From these data and the results from central kinematics (Sec.~2), there
is an emerging picture of a dichotomy in the DM distributions
of faint/disky/rotating and bright/boxy/non-rotating ellipticals.
The boxy galaxies appear DM-dominated, perhaps with strong
dissipation-driven halo contraction, 
while the disky galaxies show weak DM, indicating halos that
are less contracted or even nonexistent.
Hints of this dichotomy have appeared elsewhere
(Bertin et al. 1994; Magorrian \& Ballantyne 2001; G+01;       
Ferreras et al. 2005),
which seems opposite to what one expects in the picture of
disky and boxy galaxies formed in gas-rich and gas-poor mergers,
respectively (e.g. Naab et al. 2005).

There are a number of possible explanations for these results. One
is that spherical modeling in some of the cases could systematically
produce lower mass estimates.
Another is that the halo concentrations are lower than expected because
of baryonic physics (Mo \& Mao 2004), or
modified cosmological or dark matter models
(McGaugh et al. 2003).
Low halo concentrations have been suggested for late-type galaxies
and for ellipticals
(Rusin et al. 2003; Fukazawa et al. 2006).

Dekel et al. (2005) have shown that declining dispersion profiles
may be the natural outcome of gas-rich galaxy mergers---produced
by radial anisotropy, triaxiality,
and perhaps by contamination with younger PNe.
Whether their simulations include realistic baryonic physics
and are representative of the full cosmological picture
are open questions.
But apart from this, their merger remnants may not
resemble the real galaxies in question:
e.g., the simulated dispersion decline is partially due to a high
DM fraction in the central parts, while empirically the
disky rotators appear to have low central DM content.

Such simulations are important because even if dynamical modeling
of observed galaxies finds their mass distributions consistent
with $\Lambda$CDM halos, 
this does not show that their overall structure and kinematics are
predicted by $\Lambda$CDM.
One may consider two complementary approaches to theory testing.
The first is to fit the data with parametrized models
(e.g. mass or anisotropy parameters), and see if the results
are consistent with theory.
Such model inferences will always include uncertain assumptions
(about geometry, equilibrium, uniqueness, oversimplification).
The second approach is to ``observe'' the theory, e.g. computing from simulations
such observable quantities as $(v/\sigma)^*$ or the fundamental plane parameters. 
Special attention should be paid to the correlations between the observables.
Both approaches need large samples of galaxies, observationally
and theoretically, because of the intrinsic spread of galaxy properties.

With the advent of large samples of halo kinematics data,
we can explore suitable parametrizations (as in the second approach).
One possibility is the slope of the projected velocity dispersion profile.
With simplified $\Lambda$CDM models (assuming isotropy, $R^{1/4}$ galaxies,
no baryon contraction),
the shape of the dispersion profile measured between 3 and 8 $R_{\rm eff}$
should be fairly independent of galaxy luminosity and insensitive to
the scatter in galaxy and halo sizes.
It is thus primarily sensitive to the overall DM fraction, and
in $\Lambda$CDM the log-slope of the dispersion should be $\sim -0.1$ near
$L^*$.
The results available so far from PN kinematics do not compare favorably with
the predictions of this particular toy model
(Fig. 4, left).

\begin{figure}[t]
   \centering
   \includegraphics[width=6.195cm]{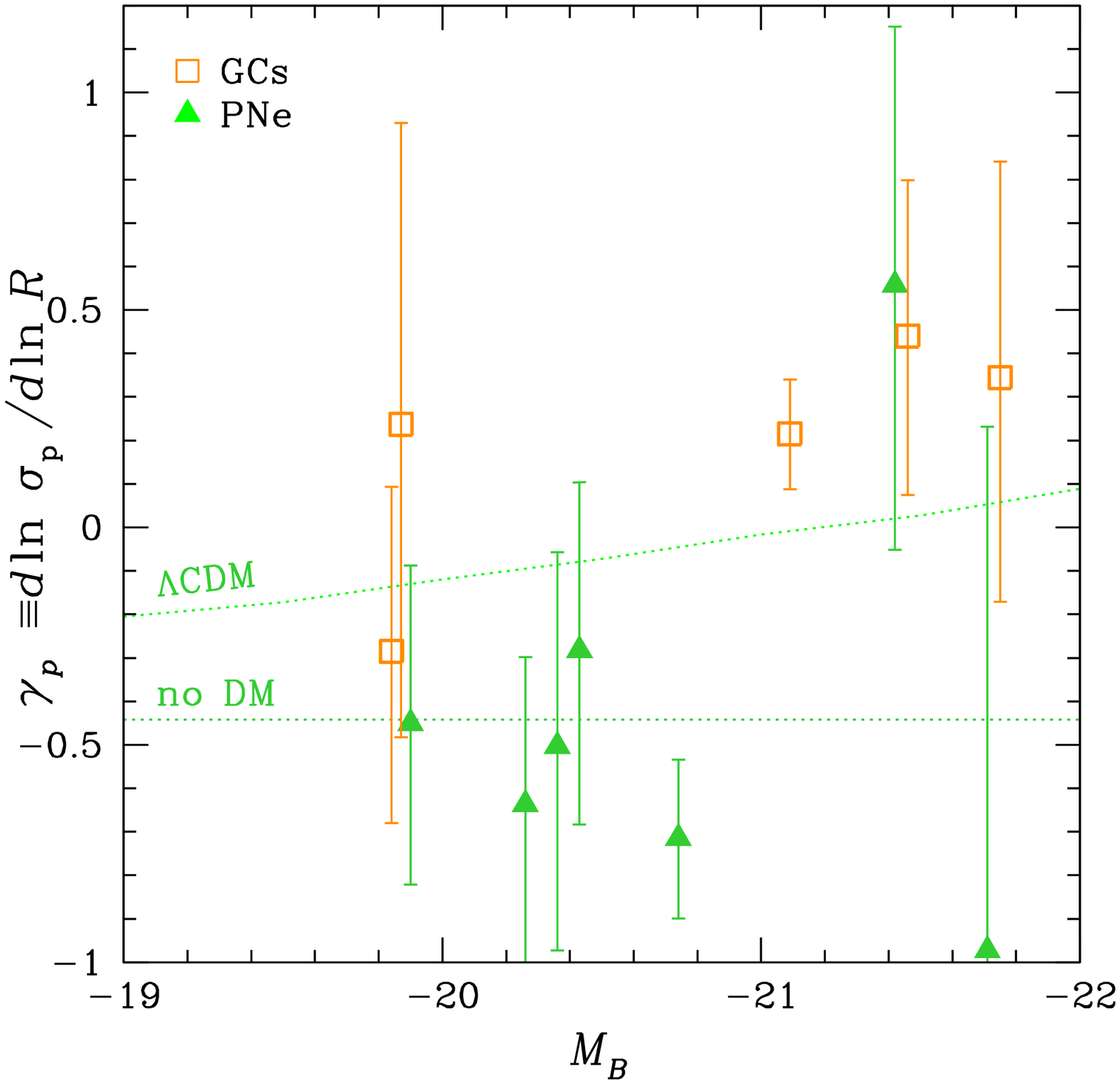}
   \includegraphics[width=6.195cm]{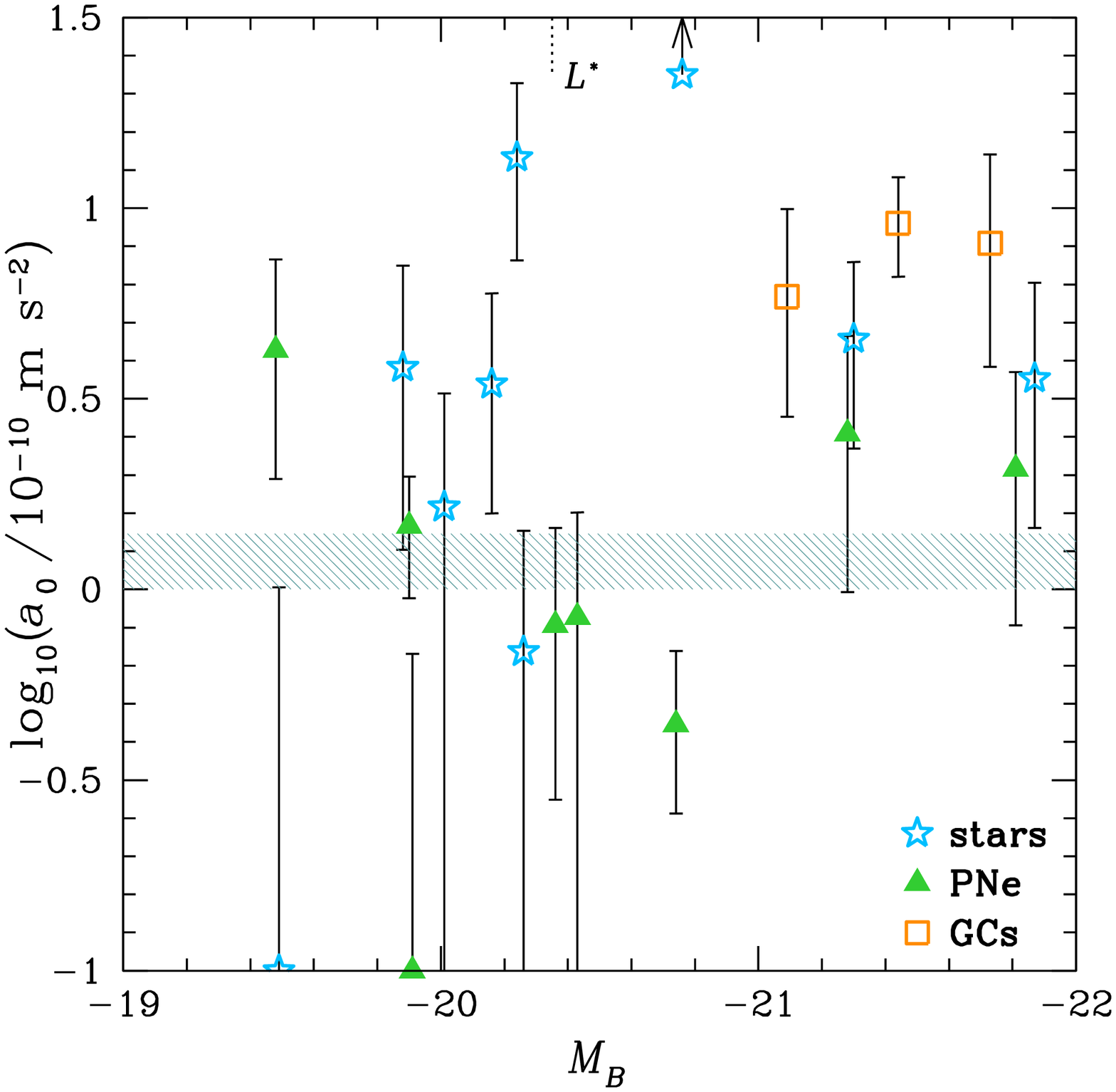}
      \caption{ 
	{\it Left:} Log slope of projected velocity dispersion profiles,
	between 3 and 8 $R_{\rm eff}$.
	Dotted lines show theoretical expectations for PNe, assuming isotropy.
	Points with error bars show observations.
	The GC data should not be compared directly to the PN models.
	{\it Right:} MOND acceleration parameter.
	The value implied by late-type galaxies is shown by a shaded region.
	Fits to halo kinematics are shown as points with error bars.
	}
       \label{figure_mafig4}
   \end{figure}

How does MOND shape up when compared to the data on ellipticals?
We have taken the same literature data and found the best-fit
acceleration parameter $a_0$ for each galaxy.
The value of $a_0$ should be universal for all types of gravitational
systems, and is found to be $\simeq (1.2 \pm 0.2) \times 10^{-10}$~m~s$^{-2}$
in late-type galaxies.
For the fainter ellipticals, this is 
consistent with most of the data
(see Fig. 4, right, and Milgrom \& Sanders 2003).
For brighter galaxies, a higher value of $a_0$ is inferred, implying
some DM is present even with MOND assumed;
this is related to the problem of DM in cluster cores (Pointecouteau \& Silk 2005).
Another modified gravity theory has also been applied to ellipticals
(Brownstein \& Moffat 2006).

Thus, while the DM content of fainter ellipticals may be a hurdle
for $\Lambda$CDM, the brighter ellipticals appear to be a roadblock for MOND.
The situation should become steadily clearer with the combined program
now underway on halo masses.







\begin{thebibliography}{}
\bibitem{}Abadi, M. G., Navarro, J. F., \& Steinmetz, M. 2005, MNRAS, astro-ph/0506659
\bibitem{}Athanassoula, E. 2005, in Planetary Nebulae as Astronomical Tools, astro-ph/0510808
\bibitem{}Baes, M., \& Dejonghe, H. 2001, ApJ, 563, L19
\bibitem{}Battaglia, G., Helmi, A., Morrison, H., Harding, P., et al. 2005, MNRAS, 364, 433
\bibitem{}Bergond, G., Zepf, S. E., Romanowsky, A. J., et al. 2006, A\&A, astro-ph/0511492
\bibitem{}Bertin, G., Bertola, F., Buson, L. M., Danziger, I. J., et al. 1994, A\&A, 292, 381
\bibitem{}Blumenthal, G. R., Faber, S. M., Flores, R., \& Primack, J. R. 1986, ApJ, 301, 27
\bibitem{}Borriello, A., Salucci, P., \& Danese, L. 2003, MNRAS, 341, 1109
\bibitem{}Brownstein, J. R., \& Moffat, J. W. 2006, ApJ, in press, astro-ph/0506370
\bibitem{}Cappellari, M., Bacon, R., Bureau, M., et al. 2005, MNRAS, astro-ph/0505042 (C+05)
\bibitem{}De Bruyne, V., De Rijcke, S., Dejonghe, H., \& Zeilinger, W. W. 2004, MNRAS, 349, 461
\bibitem{}Dekel, A., Stoehr, F., Mamon, G. A., Cox, T. J., et al. 2005, Nature, 437, 707
\bibitem{}Diemand, J., Madau, P., \& Moore, B. 2005, MNRAS, 364, 367
\bibitem{}Douglas, N. G., Arnaboldi, M., Freeman, K. C., et al. 2002, PASP, 114, 1234
\bibitem{}Ferreras, I., Saha, P., \& Williams, L. L. R. 2005, ApJ, 623, L5
\bibitem{}Ferrarese L. 2002, ApJ, 578, 90
\bibitem{}Fukazawa, Y., Betoya-Nonesa, J. G., Pu, J., et al. 2006, ApJ, in press, astro-ph/0509521
\bibitem{}Gerhard, O., Kronawitter, A., Saglia, R. P., \& Bender, R. 2001, AJ, 121, 1936 (G+01)
\bibitem{}Gonz\'{a}lez-Garc\'{i}a, A. C., \& Balcells, M. 2005, MNRAS, 357, 753
\bibitem{}Kormendy, J., \& Bender, R. 1996, ApJ, 464, L119
\bibitem{}Kronawitter A., Saglia, R. P., Gerhard, O., \& Bender, R. 2000, A\&AS, 144, 53
\bibitem{}Magorrian, J., \& Ballantyne, D. 2001, MNRAS, 322, 702
\bibitem{}Mamon, G. A., \& {\L}okas, E. L. 2005a, MNRAS, 362, 95
\bibitem{}Mamon, G. A., \& {\L}okas, E. L. 2005b, MNRAS, 363, 705
\bibitem{}Matsushita, K., Belsole, E., Finoguenov, A., \& B\"{o}hringer, H.  2002, A\&A, 386, 77
\bibitem{}McGaugh, S. S., Barker, M. K., \& de Blok, W. J. G. 2003, ApJ, 584, 566
\bibitem{}M\'{e}ndez, R. H., Riffeser, A., Kudritzki, R.-P., Matthias, M., et al. 2001, ApJ, 563, 135
\bibitem{}Merritt, D., \& Saha, P. 1993, ApJ, 409, 75
\bibitem{}Milgrom, M., \& Sanders, R. H. 2003, ApJ, 599, L25
\bibitem{}Mo, H. J., \& Mao, S. 2004, MNRAS, 353, 829
\bibitem{}Naab, T., Khochfar, S., \& Burkert, A. 2005, ApJ, submitted, astro-ph/0509667
\bibitem{}Napolitano, N. R., Capaccioli, M., et al. 2005, MNRAS, 357, 691 (N+05)
\bibitem{}Padmanabhan, N., Seljak, U., Strauss, M. A., et al. 2004, New Astron., 9, 329
\bibitem{}Peng, E. W., Ford, H. C., \& Freeman, K. C. 2004, ApJ, 602, 685
\bibitem{}Pointecouteau, E., \& Silk, J. 2005, MNRAS, 364, 654
\bibitem{}Richtler, T., Dirsch, B., Gebhardt, K., Geisler, D., Hilker, M., et al. 2004, AJ, 127, 2094
\bibitem{}Richstone, D., Gebhardt, K., Aller, M., et al. 2004, ApJL, submitted, astro-ph/0403257
\bibitem{}Romanowsky, A. J., \& Kochanek, C. S. 2001, ApJ, 553, 722
\bibitem{}Romanowsky, A. J., Douglas, N. G., Arnaboldi, M., et al. 2003, Science, 301, 1696
\bibitem{}Rusin, D., Kochanek, C. S., \& Keeton, C. R. 2003, ApJ, 595, 29
\bibitem{}S\'{a}iz, A., Dom\'{i}nguez-Tenreiro, R., \& Serna, A. 2004, ApJ, 601, L131
\bibitem{}Schneider, S. E. 1985, ApJ, 288, L33
\bibitem{}Schwarzschild, M. 1979, ApJ, 232, 236
\bibitem{}Statler, T. S., Dejonghe, H., \& Smecker-Hane, T. 1999, AJ, 117, 126
\bibitem{}Teodorescu, A. M., M\'{e}ndez, R. H., Saglia, R. P., et al. 2005, ApJ, astro-ph/0509831
\bibitem{}Thomas, J., Saglia, R. P., Bender, R., Thomas, D., et al. 2005, MNRAS, 360, 1355
\bibitem{}Trujillo, I., Burkert, A., \& Bell, E. F. 2004, ApJ, 600, L39
\bibitem{}van den Bosch, F. C., Mo, H. J., \& Yang, X. 2003, MNRAS, 345, 923
\bibitem{}van der Marel, R. P., \& Franx, M. 1993, ApJ, 407, 525
\bibitem{}Wu, X., \& Tremaine, S. 2005, ApJ, submitted, astro-ph/0508463

\end{thebibliography}
\end{document}